\documentstyle[12pt,epsf,epsfig]{article}
\newcommand{\ea}{{\it et al.}}

\newcommand{\eg}{{\it e.g. }}
\newcommand{\mrm}[1]{\mbox{\rm #1}}
\newcommand{\beq}{\begin{equation}}
\newcommand{\eeq}{\end{equation}}
\newcommand{\bea}{\begin{eqnarray}}
\newcommand{\eea}{\end{eqnarray}}
\newcommand{\rfn}[1]{(\ref{#1})}

\newcommand{\nn}{\nonumber}

\newcommand{\prl}[1]{{ Phys. Rev. Lett. }{\bf #1}}

\def\lsim{\mathrel{\vcenter{\hbox{$<$}\nointerlineskip\hbox{$\sim$}}}}
\def\gsim{\mathrel{\vcenter{\hbox{$>$}\nointerlineskip\hbox{$\sim$}}}}

\def\m12{m_{1\!/2}}

\begin{document}
\begin{titlepage}
\pagestyle{empty}
\baselineskip=21pt
\rightline{CERN--TH/2001-311}     
\rightline{KEK-TH-791}
\rightline{DPNU-01-32}
\vskip 0.25in
\begin{center}
{\large{\bf 
Lepton Electric Dipole Moments in Non-Degenerate 
Supersymmetric Seesaw Models
}}
\end{center}
\begin{center}
\vskip 0.25in
{{\bf John Ellis}$^1$,
{\bf Junji Hisano}$^{2}$,
{\bf Martti Raidal}$^{1,3}$ and
{\bf Yasuhiro Shimizu}$^4$
\vskip 0.15in
{\it
$^1${Theory Division, CERN, CH-1211 Geneva 23, Switzerland}\\
$^2${Theory Group, KEK, Oho 1-1, Tsukuba, Ibaraki 305-0801, Japan}\\
$^3${National Institute of Chemical Physics and Biophysics, 
Tallinn 10143, Estonia}\\ 
$^4${Department of Physics, Nagoya University, Nagoya 464-8692, Japan}\\
}}
\vskip 0.25in
{\bf Abstract}
\end{center}
\baselineskip=18pt \noindent

In the context of supersymmetric seesaw models of neutrino masses with
non-degenerate heavy neutrinos, we show that Dirac Yukawa interactions
$N^c_i (Y_\nu)_{ij} L_j H_2$ induce large threshold corrections to the
slepton soft masses via renormalization. While still yielding rates for
lepton-flavour-violating processes below the experimental bounds, these
contributions may increase the muon and electron electric dipole moments
$d_\mu$ and $d_e$ by several orders of magnitude. In the leading
logarithmic approximation, this is due to three additional physical phases
in $Y_\nu,$ one of which also contributes to leptogenesis. The naive
relation $d_\mu/d_e\approx -m_\mu/m_e$ is violated strongly in the case of
successful phenomenological textures for $Y_\nu$, and the values of
$d_\mu$ and/or $d_e$ may be within the range of interest for the future
experiments.

\vfill
\leftline{CERN--TH/2001-311}
\leftline{November 2001}
\end{titlepage}
\baselineskip=18pt


Understanding particle masses and mixings, is currently one of the most
important issues in high-energy physics, together with the associated
phenomenology. On the experimental side, this is manifested by many
ongoing and planned neutrino oscillation experiments and searches for
lepton flavour violation (LFV), as well as by the continuous improvement
in the measurements of mixing parameters in the quark sector, most
recently of CP violation at the $B$ factories. On the theory side, the
importance is manifested by enormous efforts in the interpretation of
these measurements using various ideas beyond the Standard Model. The
smallness of neutrino masses is commonly explained by the seesaw
mechanism~\cite{seesaw}, which introduces heavy singlet (right-handed)  
neutrinos $N_i$ with masses somewhat below the unification scale.
Requiring the corresponding hierarchy of mass scales to be natural
motivates supersymmetrizing the theory to cancel the quadratic divergences
in the Higgs boson masses. The absence of large supersymmetric
contributions to flavour- and CP-violating processes~\cite{masiero}
suggests that the soft supersymmetry-breaking terms are real and universal
at the unification scale. In this case, the flavour and CP violation in
the slepton sector are entirely induced by the renormalization effects of
neutrino coupling parameters~\cite{bm}.

For simplicity, in almost previous studies of renormalization-induced
LFV and CP violation in the lepton sector, the heavy singlet neutrino
masses have generally not been distinguished, and hence they have been
integrated out at the same scale.\footnote{
In Ref.~\cite{Blazek:2001zm}, $\tau\rightarrow \mu \gamma$ is
evaluated in the supersymmetric see-saw model, including
non-degeneracy of the right-handed neutrino masses.
} 
In this case, observable rates for LFV processes such as $\mu\to
e\gamma$, $\mu$-$e$ conversion in nuclei, $\mu\to eee$ and $\tau \to 3
\ell$~\cite{review,h1,Okada} can be generated in the minimal
supersymmetric seesaw model, but the electric dipole moments (EDMs) of
charged leptons are many orders of magnitude below the sensitivities
of future experiments~\cite{strumia}. The latter suppression is due to
the fact that, when the heavy singlet neutrinos are degenerate, only
one of the physical CP phases in the the minimal supersymmetric seesaw
model contributes to EDMs, and the CP violation in the lepton sector
is entirely connected to the flavour violation, just as in the quark
sector in the Standard Model.

However, there is no reason to believe that the heavy neutrino masses
$M_{N_i}$ are exactly degenerate.  Indeed, the known hierarchies in the
quark and lepton masses suggest the opposite. Moreover, models of Yukawa
textures~\cite{ross} that successfully explain the quark, charged lepton
and light neutrino parameters in fact predict very hierarchical
right-handed neutrino masses~\cite{lms,ross2,bw}. In addition,
leptogenesis~\cite{lepto}, the only known manifestation of leptonic CP
violation so far, requires non-degenerate singlet heavy neutrino masses,
and very hierarchical $M_{N_i}$ are preferred in specific models such as a
supersymmetric $SO(10)$ grand unified theory~\cite{so10lepto}.

In this letter we investigate the effects of non-degeneracy of the
right-handed neutrinos on the LFV and CP-violating observables in the
minimal supersymmetric seesaw model. We show that, if the heavy neutrinos
are integrated out at different scales, the threshold corrections to the
soft slepton mass terms proportional to $\ln (M_{N_i}/M_{N}),$ where $M_N$
is a geometric mean of $M_{N_i},$ may strongly affect charged-lepton EDMs,
$Br(\mu\to e\gamma)$ and to a lesser extent also $Br(\tau\to\mu\gamma).$
In the leading logarithmic approximation, these non-degeneracy
contributions introduce into the EDMs dependences on three additional
phases, one of which contributes to leptogenesis, that do not renormalize
soft slepton mass terms if the heavy singlet neutrinos are degenerate. The
charged-lepton EDMs may be enhanced by {\it several orders of magnitude}
over the degenerate neutrino case, bringing them potentially within the
range of interest for foreseeable future
experiments~\cite{mnbuedm,PRISM,nf,edmnew}.

We illustrate our general arguments with specific numerical examples using
phenomenologically successful symmetric 
neutrino mass textures~\cite{ross2,bw}.
Taking into account the present bounds on $Br(\mu\to e\gamma)$ and
$Br(\tau\to\mu\gamma),$ as well as the recent measurement of $(g_\mu-2)$
and direct bounds on sparticle masses, the textures considered
may yield values of the muon EDM $d_\mu$
exceeding $\sim 10^{-26}$~e~cm, and the electron EDM $d_e$ may
approach the level of $\sim 10^{-31}$~e~cm. The naive relation
$d_\mu/d_e\approx -m_\mu/m_e$, which holds well in the degenerate
heavy-neutrino case, is badly violated for non-degenerate heavy 
neutrinos.\footnote{
It is known that the relation $d_\mu/d_e\approx -m_\mu/m_e$ does not
hold in the SUSY SO(10) GUT \cite{Barbieri:1995tw}, the non-minimal
SUSY SUSY SU(5) GUTs \cite{Arkani-Hamed:1995fs}, and the
supersymmetric left-right models \cite{Babu:2000dq}.}
Since the PRISM experiment~\cite{PRISM} and neutrino factory stopped-muon
experiments~\cite{nf} aim at a sensitivity $d_\mu\sim 10^{-26}$~e~cm, and
a newly proposed technology might allow a sensitivity to $d_e\sim
10^{-33}$~e~cm~\cite{edmnew}, our results suggest that the minimal
supersymmetric seesaw model may be testable in these experiments.
Combining the lepton EDM measurements with the CP- and T-violating
observables~\cite{cpt} in neutrino oscillations using superbeams or in
neutrino factory long-baseline experiments, with the possible measurement
of T-odd asymmetries in $\mu\to eee$ \cite{ehlr}, and with CP violation in
slepton oscillations at colliders \cite{acfh} and in $\beta\beta_{0\nu}$
decays, one may be able to test the leptogenesis mechanism using a
programme of low-energy experiments. In addition, if $d_\mu$ is large,
$Br(\tau\to\mu\gamma)$ must be close to the present experimental bound,
making possible its detection in $B$-factory or LHC experiments.

We start our studies by considering the following leptonic superpotential, 
which implements the seesaw mechanism in a minimal way:
\begin{eqnarray}
\label{w}
W = N^{c}_i (Y_\nu)_{ij} L_j H_2
  -  E^{c}_i (Y_e)_{ij}  L_j H_1 
  + \frac{1}{2}{N^c}_i {\cal M}_{ij} N^c_j + \mu H_2 H_1 \,,
\label{MseesawM}
\end{eqnarray}
where the indices $i,j$ run over three generations and ${\cal M}_{ij}$
is the heavy singlet-neutrino mass matrix. One can always work in  a basis 
where the charged leptons and
the heavy neutrinos both have real and diagonal mass matrices:
\begin{equation}
(Y_e)_{ij} = {Y}^D_{e_i} \delta_{ij}\,, \;
{\cal M}_{ij} = {\cal M}^D_i \delta_{ij}=\mrm{diag}(M_{N_1},M_{N_2},M_{N_3})\,.
\end{equation}
The matrix $Y_\nu$ contains six physical phases \cite{valle} and  can be 
parametrised as \cite{ehlr}
\bea
(Y_\nu)_{ij} = Z^\star_{ik} {Y}^D_{\nu_k} X^\dagger_{kj},
\label{Y}
\eea
where $X$ is the analogue of the quark CKM matrix in the lepton sector
and has only one physical phase, and $Z= P_1 \overline{Z} P_2,$
where $\overline{Z}$ is a CKM-type matrix with three real mixing
angles and one physical phase, and
$P_{1,2}=\mrm{diag}(e^{i\theta_{1,3}}, e^{i\theta_{2,4}}, 1 ).$

The soft supersymmetry-breaking terms in the leptonic sector are
\begin{eqnarray}
-{\cal L}_{\rm soft}&=& 
  \tilde{L}_{i}^{\dagger} (m_{\tilde L}^2)_{ij}\tilde{L}_{j} 
+ \tilde{ E}_{i}^{c} (m_{\tilde E}^2)_{ij}\tilde{E}_{j}^{c\ast}  
+ \tilde{N}_{i}^{c} (m_{{\tilde N}}^2)_{ij} \tilde{N}^{c\ast}_{j}  
\nonumber\\
&& 
 + \left(\tilde{N}^{c}_{i} (A_{N})_{ij} \tilde{L}_{j} H_2
  -\tilde{E}^{c}_{i} (A_{e})_{ij}\tilde{L}_{j} H_1
 +\frac12 \tilde{N}_{i}^{c} (B_{N})_{ij}\tilde{N}^{c\ast}_{j}  
\right. \nn \\
&& \left.
 +\frac12 M_1 \tilde{B} \tilde{B}
 + \frac12 M_2 \tilde{W}^a \tilde{W}^a 
 + \frac12 M_3 \tilde{g}^a \tilde{g}^a
   + h.c.\right) .
\end{eqnarray}
Motivated by experimental upper limits on supersymmetric contributions to 
LFV and CP-violating effects, we assume 
universal boundary conditions at the GUT scale $M_{GUT} \sim 2\times
10^{16}$ GeV:
\bea
& 
(m^2_{\tilde E})_{ij}=(m^2_{\tilde L})_{ij}=(m^2_{\tilde N})_{ij}=
m_0^2 {\bf 1},
&\nn \\
& m^2_{H_1}=m^2_{ H_2}=m_0\,, & \nn\\
&(A_e)_{ij}=A_0 (Y_e)_{ij}\,,(A_\nu)_{ij}=A_0 (Y_\nu)_{ij}\,,   & \nn\\
& M_1=M_2=M_3=m_{1/2}\,.&
\label{mssmbound}
\eea
In this case, renormalization induces
sensitivity to the neutrino Yukawa couplings ${ Y}_\nu$ in the soft
supersymmetry-breaking parameters. 
These may induce measurable  LFV decays and CP observables, such as
EDMs, as we demonstrate by studying approximate analytical
solutions to the renormalization-group equations (RGE).

If the heavy neutrinos are exactly degenerate with a common mass $M_N$,
the flavour-dependent parts of the induced soft supersymmetry-breaking
terms are given in the leading-logarithmic approximation by
\bea
\label{llrge} 
\left(\delta m_{\tilde L}^2\right)_{ij} & \approx & 
-\frac{1}{8\pi^2}(3m_0^2 + A_0^2)
({ Y_\nu^\dagger}{ Y_\nu}  +{ Y_e^\dagger}{ Y_e})_{ij}
\log\frac{M_{GUT}}{M_N}\ , 
\nn \\
\left(\delta m_{\tilde E}^2\right)_{ij} & \approx & 
-\frac{1}{4\pi^2}(3m_0^2 + A_0^2)
({ Y_e}{ Y_e^\dagger})_{ji}
\log\frac{M_{GUT}}{M_N}\ , 
\nn \\
(\delta {A_e})_{ij}& \approx &  
- \frac{1}{8\pi^2} A_0 Y_{e_i} (
3 { Y_e^\dagger}{Y_e}
+ { Y_\nu^\dagger}{Y_\nu})_{ij}
\log\frac{M_{GUT}}{M_N} .
\eea
Here, the Yukawa coupling constants are given at $M_N$, and then
${Y_e}$ is diagonal. This means that $m^2_{\tilde E}$ remains diagonal
in this approximation. Below $M_N$, the heavy neutrinos decouple, and
the renormalization-group running is given entirely in terms of the
MSSM particles and couplings, and is independent of $Y_\nu$. 
It is important to notice that the only combination of neutrino Yukawa
couplings entering (\ref{llrge})  is $Y_\nu^{\dagger} Y_\nu$. 
It is straightforward to see from (\ref{Y}) that
\begin{equation}
\label{y+y1}
Y_\nu^{\dagger} Y_{\nu} = X (Y_\nu{^D})^2 X^{\dagger} ,
\end{equation}
and CP violation in the charged LFV processes
arises only from the single physical phase in the diagonalizing matrix $X$.
This implies that CP-violating phases are
induced only in the off-diagonal elements of $(m_{\tilde L}^2)_{ij}$
and $({A_e})_{ij}$, and further indicates that lepton-flavour-conserving
but CP-violating observables such as the electric dipole
moments of charged leptons are naturally 
suppressed~\cite{strumia,ehlr,khalil}~\footnote{Similar arguments 
hold also for the renormalization-induced neutron EDM in supersymmetric 
models~\cite{nedm}.}.
Exactly as in the Standard Model for quarks, three generations are 
necessary for physical 
CP violation.

On the other hand, if the heavy neutrinos are non-degenerate: $M_{N_i} \ne
M_{N_j},$
one obtains additional corrections from the RGEs:
\bea
(\delta m^2_{\tilde L})_{ij} \to (\delta m^2_{\tilde L})_{ij} +
\left({\tilde \delta} m_{\tilde L}^2 \right)_{ij}, 
\eea
where
\bea
\label{nondeg}
\left({\tilde \delta}
m_{\tilde L}^2\right)_{ij} \approx
-\frac{1}{8\pi^2}(3m_0^2 + A_0^2)
({ Y_\nu^\dagger} L { Y_\nu})_{ij} \;\; : \;\; 
L \equiv \log\frac{M_{N}}{M_{N_i}} \delta_{ij}\ .
\eea
Here $M_{N}$ should now be interpreted as the geometric mean of the heavy
neutrino mass eigenvalues $M_{N_i}$. The potentially large logarithm
$\log\frac{M_{N}}{M_{N_i}}$  arises from the distinct thresholds of the
heavy neutrinos. Its most important effect concerns CP violation.
Namely, according to \rfn{Y},
the first term in (\ref{nondeg}) contains the matrix factor
\begin{equation}
Y^\dagger L Y = X Y^D P_2 {\overline Z}^T L {\overline Z}^*
P_2^* Y^D X^\dagger,
\label{newlabel}
\end{equation}
which induces dependences on the phases in 
${\overline Z}$ and $ P_2$. In the
three-generation case, there are two independent entries in the traceless
diagonal matrix $L$, so the renormalization induces in principle
dependences on two new combinations of these phases, as well as the single
phase in ${Y_\nu^\dagger}{Y_\nu}$~\footnote{Moreover, CP 
violation could in principle now be observable even in a two-generation 
model.}.

What is the physical interpretation of these new phases?
At present, our only experimental knowledge of CP violation in the lepton
sector comes from the baryon asymmetry of the Universe, assuming
that this originates from leptogenesis. The $L$ asymmetry $\varepsilon^i$ 
in the decay of an individual heavy-neutrino species $N^c_i$ is given in 
the supersymmetric case~\cite{vissani} by
\begin{eqnarray}
\varepsilon^i &=& -\frac{1}{8 \pi} \sum_{l} 
\frac{ \mbox{Im}\Big[
\left( { Y_\nu}{ Y_\nu}^\dagger  \right)^{li}
\left( { Y_\nu}{ Y_\nu}^\dagger \right)^{li}
\Big]}
{ \sum_{j} |{ Y_\nu}^{ij}|^2 }
\nn \\
& &
\sqrt{x_l} \Big[  \mbox{Log} (1+1/x_l) +  \frac{2}{(x_l-1)}\Big] , 
\label{eps}
\end{eqnarray}
where $x_l \equiv (M_{N_l} / M_{N_i})^2$, and both vertex and 
self-energy
type loop diagrams are taken into account. This $L$ asymmetry is
converted into the observed baryon asymmetry by sphalerons acting
before the electroweak phase transition. It is clear from (\ref{eps})
that the generated asymmetry depends only on the three phases in
\begin {equation}
\label{yy+1}
Y_\nu Y_\nu^\dagger 
 = P_1^\star \overline{Z}^\star (Y_\nu^D)^2 \overline{Z}^T P_1\, ,
\end{equation}
namely the single phase in $\overline{Z}$ and the two
phases $\theta_{1,2}$ in $P_1.$ We note that, according to \rfn{newlabel},
the CP-violating observables at low energies depend on the leptogenesis
phase in $\overline{Z}$ and on the two phases in $P_2$. We have explicitly 
checked with our numerical programs that the dependence of the EDMs on the 
phases in $P_1$ is negligible.

Our main objective in this letter is to show that the muon and the
electron EDMs are enhanced if the heavy neutrinos are non-degenerate and
depend strongly on the leptogenesis phase as discussed above. We now
illustrate this with approximate analytical expressions in a
mass-insertion approximation.


Since the EDMs are flavour-diagonal observables and
induced by the radiative corrections to the soft supersymmetry-breaking
terms, the leading contributions  are
proportional to ${\cal O}(\log^3\frac{M_{GUT}}{M_N})$ and  to
${\cal O}(\log^2\frac{M_{GUT}}{M_N}\log\frac{M_{N}}{M_{N_i}}).$
The former contributions are present even if the heavy neutrinos are
degenerate, whilst the latter require non-degenerate heavy neutrinos.
In order to evaluate the contribution to the EDMs 
at ${\cal O}(\log^3\frac{M_{GUT}}{M_N})$, we need the corrections to the 
soft supersymmetry-breaking terms at ${\cal O}(\log^2 
\frac{M_{GUT}}{M_N})$, which are
\bea
\label{llrge2} 
\left(\delta^{(2)} m_{\tilde L}^2\right)_{ij} & \approx & 
\frac{4}{(4\pi)^2} A_0^2
(3{ Y_\nu^\dagger}{ Y_\nu} { Y_\nu^\dagger}{ Y_\nu} 
 +
3  { Y_e^\dagger}{ Y_e}{ Y_e^\dagger}{ Y_e}
+\{{ Y_e^\dagger}{ Y_e},~{ Y_\nu^\dagger}{ Y_\nu}\})_{ij}
\log^2\frac{M_{GUT}}{M_N}\ , 
\nn \\
\left(\delta^{(2)} m_{\tilde E}^2\right)_{ij} & \approx & 
\frac{8}{(4\pi)^2} A_0^2
(3 { Y_e}{ Y_e^\dagger}{ Y_e} { Y_e^\dagger}
+{ Y_e}{ Y_\nu^\dagger}{ Y_\nu}{ Y_e^\dagger} )_{ji}
\log^2\frac{M_{GUT}}{M_N}\ , 
\nn \\
(\delta^{(2)} {A_e})_{ij}& \approx &  0 .
\eea
From these equations and from (\ref{llrge}), 
non-vanishing contributions to
EDMs arise from the combinations 
$\mrm{Im} [\delta A_e \delta A_e^\dagger \delta A_e ]_{ii},$ 
$\mrm{Im}[\delta {A_e} {Y_e^\dagger}{Y_e} 
\delta^{(2)} m_{\tilde L}^2]_{ii}$
of order ${\cal O}(\log^3 \frac{M_{GUT}}{M_N})$.
They are proportional to Jarlskog invariants 
\begin{eqnarray}
J_{\nu_i}&=&\mbox{Im}[
Y_e 
Y_\nu^\dagger Y_\nu
Y_e^\dagger Y_e
Y_\nu^\dagger Y_\nu
Y_\nu^\dagger Y_\nu
]_{ii},
\end{eqnarray}
which consist of only the Yukawa couplings $Y_\nu$ and $Y_e$,
and depend on only one phase in $X$.
On the other hand, if the heavy neutrinos are not degenerate in mass, 
there are corrections of ${\cal O}(\log \frac{M_{GUT}}{M_N}\log
 \frac{M_N}{M_{N_i}})$ of the form
\bea
\label{llrge3} 
\left(\tilde{\delta}^{(2)} m_{\tilde L}^2\right)_{ij} & \approx & 
\frac{18}{(4\pi)^4} (m_0^2 +A_0^2)
\{
{ Y_\nu^\dagger}L { Y_\nu}, 
{Y_\nu^\dagger}{ Y_\nu} 
\}
\log\frac{M_{GUT}}{M_N}\ , 
\nn \\
\left(\tilde{\delta}^{(2)} m_{\tilde E}^2\right)_{ij} & \approx & 
0,
\nn \\
\left(\tilde{\delta}^{(2)} {A_e}\right)_{ij}& \approx &  
\frac{1}{(4\pi)^4} A_0 Y_e
(
11 \{
{ Y_\nu^\dagger}L { Y_\nu}, 
{Y_\nu^\dagger}{ Y_\nu} 
\}
+
7 [
{ Y_\nu^\dagger} L { Y_\nu}, 
{Y_\nu^\dagger}{ Y_\nu} 
]
)_{ij}
\log\frac{M_{GUT}}{M_N}\ , 
\eea
where we have neglected terms with the ${Y_e^\dagger}{ Y_e}$ factors. 
The crucial point is that the second term in $\tilde{\delta}^{(2)} {A_e}$ 
can have imaginary parts in its diagonal components, and thus can 
contribute to the electric dipole moment~\footnote{
Whilst the combination $(\tilde{\delta}A_e \delta m_{\tilde
L}^2)_{ii}$ has an imaginary part, it does not contribute to the electric
dipole moments, since $\mrm{Im}[(\delta A_e + \tilde{\delta}A_e)
(\delta m_{\tilde L}^2+\tilde{\delta} m_{\tilde L}^2)]_{ii}=0$. 
}.
Therefore, the explicit comparison of (\ref{llrge2}) and (\ref{llrge3})
leads us to the following conclusions:
\begin{enumerate}
\item[(i)]
While the induced lepton EDMs depend on the single physical phase in  
${Y_\nu^\dagger}{ Y_\nu}$ if the heavy neutrinos are degenerate,
for the non-degenerate case the EDMs depend on three combinations of 
phases in ${Y_\nu^\dagger}{ Y_\nu}$ and ${\overline Z} P_2,$ including
the leptogenesis phase. In the latter case CP violation can occur even 
if there are two generations of particles. 

\item[(ii)]
The EDMs depend very strongly
on the non-degeneracy of heavy neutrinos. In particular, a
step-function-like enhancement of (some) EDMs is expected when going from the 
degenerate to the non-degenerate case. This is associated with 
the disappearance of the suppression of CP invariants by the
light generations.

\item[(iii)]
The EDMs depend strongly on $A_0.$ 
\end{enumerate}
With these observations in mind, we proceed to exact numerical
calculations.

In SUSY models the LFV decays $l_i\to l_j\gamma$ and the lepton
EDMs are both generated by one-loop Feynman diagrams with 
neutralinos/charged sleptons and charginos/sneutrinos running in
the loop. For the LFV decays, the effective Lagrangian reads 
\begin{eqnarray}
{\cal L} &=& -\frac{e}{2} \left\{  
        {m_{i }}{A_L}\overline{l}_{jR}
        {{\sigma }^{\mu \nu}{l_{iL}}{F_{\mu \nu}}}
       + {m_{i }}{A_R}\overline{l}_{jL}
        {{\sigma }^{\mu \nu}{l_{iR}}{F_{\mu \nu}}}\right\},
\label{eq:effective}
\end{eqnarray}
and the decay rates are given by
\bea
\Gamma( l_i\to l_j\gamma )=\frac{e^2}{16 \pi} m^5_i
\left(|A_L|^2+|A_R|^2 \right)\,.
\eea
The explicit expressions for $A_{L,R}$  in terms of the
supersymmetric charged- and 
neutral-current parameters can be found in \cite{h1,Okada},
and we do not present them here.
Similarly, the electric dipole moment of a lepton $l$ is defined as 
the coefficient $d_l$ of the effective interaction
\bea
{\cal L} = 
-\frac{i}{2} d_l\,\bar l\, \sigma_{\mu\nu}\gamma_5\, l\,
F^{\mu\nu},
\label{edm_eff}
\eea
and can be expressed as
\bea
d_l=d^{\chi^+}_l+d^{\chi^0}_l\,,
\eea
where \cite{nath,khalil}
\bea
\label{dl+}
 d_l^{\chi^+}&=&-\frac{e}{(4\pi)^2} 
      \sum_{A=1}^{2}\sum_{X=1}^{3} 
 {\rm Im}(C^L_{lAX} C^{R*}_{lAX})\;   
 {m_{\chi^+_A}\over {m_{\tilde{\nu}_X}^2}}
   {\rm A}\biggl( \frac{m_{\chi^+_A}^2}{m_{\tilde{\nu}_X}^2} \biggr)\;,
\label{dlc}
\\ 
\label{dl0}
 d_l^{\chi^0 }&=&-\frac{e}{(4\pi)^2}
   \sum_{A=1}^{4}\sum_{X=1}^{6} 
{\rm Im}( N^L_{lAX} N^{R*}_{lAX}  )
               \frac{m_{\chi^0_A}}{M_{\tilde{l}_X}^2}\;
{\rm B}\biggl( \frac{m_{\chi^0_A}^2}{M_{\tilde{l}_X}^2}\biggr) \;,
\eea
and the loop functions are given by
\bea
&& A(x)=\frac{1}{2(1-x)^2}\biggl(3-x+\frac{2\log x}{1-x}\biggr) \;, 
\nonumber\\
&& B(x)=\frac{1}{2(1-x)^2}\biggl(1+x+\frac{2r\log x}{1-x}\biggr) \; . \nn
\eea
Here the relevant chargino and neutralino couplings $C^{L,R}$ and $N^{L,R}$ 
can be found in \cite{h1,Okada}.

However, in the present scenario where the CP-violating phases
are generated only in the off-diagonal elements of the slepton 
mass matrix, it may turn out that \rfn{dlc},\rfn{dl0} are not
useful in numerical calculations. Because of 
finite computer accuracy, we find that orders of magnitude 
numerical errors in evaluating the EDMs may occur if the off-diagonal
CP phases are small. Therefore we present here the expressions for 
EDMs in the mass-insertion approximation. These allow precise and
reliable calculation of EDMs even for very small CP-violating effects.

When the CP-violating phases are attributed to the off-diagonal components
in the slepton mass matrix, the Bino-like neutralino diagrams are 
the dominant ones. Mass insertions to order $n$ then yield
the formula:
\begin{eqnarray}
\frac{d_l}{e}
&=&
\frac{g_Y^2}{(4 \pi)^2}
\sum_{A} O_{A1} M_{\tilde{\chi}_A} (O_{A1}+O_{A1}/\tan\theta_W)
\nonumber\\
&&
\sum_n \sum_{l_1,\cdots\l_{n-1}}
{\mbox{Im}}\left[
\Delta(m^2_{\tilde{l}})_{l',l_1}
\Delta(m^2_{\tilde{l}})_{l_1 l_2}
\cdots
\Delta(m^2_{\tilde{l}})_{l_{n-1}l}
\right]
\nonumber\\
&&f_n(
(m^2_{\tilde{l}})_{l'l'},
(m^2_{\tilde{l}})_{l_1l_1},
\cdots,
(m^2_{\tilde{l}})_{l_{n-1}l_{n-1}},
(m^2_{\tilde{l}})_{ll};
|M_{\tilde{\chi}_A}|^2
) \,,
\end{eqnarray}
where $O$ is the neutralino mixing matrix,
$l'=l+3,$ $l_1,...,l_{n-1}=1,...,6,$ $A=1,...,4$ and 
$\Delta(m^2_{\tilde{l}})_{l_1,l_2}
= (m^2_{\tilde{l}})_{l_1,l_2} (i\ne j)$. 
The mass functions $f_n$ are given by the finite differences
\begin{eqnarray}
f_n(m^2_1,\cdots,m^2_{n+1};|M|^2)
&=&
\frac{1}{m^2_1-m^2_2}
\left(
f_{n-1}(m^2_1,m^2_3,\cdots,m^2_{n+1}; |M|^2)
\right.
\nonumber\\
&&
\left.-
f_{n-1}(m^2_2,m^2_3,\cdots,m^2_{n+1}; |M|^2)
\right) \,,
\label{fd}
\end{eqnarray}
where the zeroth  function is $f_0(m^2;|M|^2) = 1/|M|^2 f(m^2/|M|^2)$ 
and 
\begin{eqnarray}
f(x)&=& \frac{1}{2(1-x)^3}
\left( 1-x^2+2x \log x \right) \,.
\end{eqnarray}
For a precise evaluation of EDMs even in a case of degenerate heavy
neutrinos, sixth-order terms are needed, since they yield $\mrm{Im}[A_e
m_{\tilde L}^2 A_e^\dagger A_e m_{\tilde L}^2 m_{\tilde L}^2]$. 
Comparison with the exact calculation also shows that this order is also
sufficient for an accurate result.

The current experimental bounds on the LFV decays are
${Br}(\mu\to e\gamma)\lsim 1.2\times 10^{-11}$ \cite{mega} and 
${Br}(\tau\to\mu\gamma)\lsim 1.1\times 10^{-6}$ \cite{cleo}. 
An experiment with the sensitivity
${Br}(\mu\to e\gamma)\sim 10^{-14}$ is proposed at PSI \cite{PSI}
and the stopped-muon experiments at neutrino factories will reach 
${Br}(\mu\to e\gamma) \sim 10^{-15}$ \cite{nf}.
The bound on ${Br}(\tau\to\mu\gamma)$ will be improved at LHC and B-factories
by an order of magnitude.

For EDMs the bounds are $d_e<4.3\times 10^{-27}$ e cm for the
electron \cite{eedm}, $d_\mu = (3.7 \pm 3.4) \times 10^{-19}$ e cm 
for the muon \cite{muedm}, and
$|d_\tau| < 3.1
\times 10^{-16}$ e cm for the $\tau$ \cite{tauedm}. An experiment has 
been proposed at
BNL that could improve the sensitivity to $d_\mu$ down to $d_\mu \sim
10^{-24}$ e cm \cite{mnbuedm}, and PRISM \cite{PRISM}  
and neutrino factory \cite{nf} experiments 
aim at sensitivities $d_\mu \sim 5\times 10^{-26}$ e cm. 
Recently it has been proposed \cite{edmnew} 
that using new technology one could
improve the upper bound on the electron EDM by six orders of
magnitude and reach $d_e \sim 10^{-33}$ e cm.
We shall show that
these expected sensitivities will allow one to test the minimal
supersymmetric seesaw model.

Our calculational procedure is to fix the gauge and the quark and
charged-lepton Yukawa couplings at $M_Z$ and run them to the scale $M_{N_1}$
using the one-loop MSSM RGEs. Above $M_{N_1}$ we use the RGE-s for
the SUSY seesaw model~\footnote{Notice that in addition to notational
differences in the superpotential and soft mass terms, also the sign 
of the Yukawa Lagrangian varies in the literature. This affects also
the RGEs, as may be seen by comparing~\cite{h1} with~\cite{mv}.
The plots here are made in the convention of~\cite{mv}.} 
and integrate--in the heavy neutrinos
each at its own mass scale. At the GUT scale we choose universal
boundary conditions for the soft masses and run all the parameters 
down again, integrating each $N_i$ out at its
own mass scale. In this way we get at the $M_Z$ scale the $3\times 3$ soft 
mass matrices which include the heavy neutrino threshold effects.
These are used to calculate the LFV processes and the EDMs.

In order to implement this procedure, we have to specify the neutrino
Yukawa matrix $Y_\nu$ which is the only source of LFV and CP violation in
our model.  We follow recent works on
the Yukawa textures~\cite{ross,ross2,bw},
and choose a generic form of $Y_\nu$ of the form
\bea
Y_\nu= Y_0
\left(\begin{array}{ccc}
0 & c \,\varepsilon_\nu^3 & d\, \varepsilon_\nu^3 \\
c\, \varepsilon_\nu^3 & a \,\varepsilon_\nu^2 & b\, \varepsilon_\nu^2 \\
d\, \varepsilon_\nu^3 &  b\, \varepsilon_\nu^2 & e^{i\psi}
\end{array} \right) \, .
\label{texture}
\eea
Here $\varepsilon_\nu$ is a hierarchy parameter which determines the 
flavour mixings, $a,$ $b,$ $c$ and $d$ are complex numbers, 
and $Y_0$ is an overall scale factor.  
These textures are constructed to satisfy the measured neutrino
mass and mixing parameters via the seesaw mechanism, and in different 
models the parameters in \rfn{texture} may have different values.
For example, in the SO(10) GUT-motivated model \cite{bw} $d=0$ has been
taken for simplicity, while the model \cite{ross2} with SU(3) family symmetry
predicts $a=b$ and $d=c.$ The parameter $\varepsilon_\nu$ is
in principle a free parameter to be fixed from neutrino phenomenology.
These models predict also very hierarchical heavy neutrinos. 
Following~\cite{bw}, we fix their hierarchy to be
\bea
M_{N_1}\; : \;M_{N_2}\; : \;M_{N_3}\; = \; 
\varepsilon_N^6\; : \;\varepsilon_N^4\; : \;1 \,,
\eea
but we treat $\varepsilon_N$ as an independent phenomenological parameter. 
This allows us roughly to cover the phenomenology of different models,
as we discuss below.

We emphasise that the details of the effective light-neutrino parameters 
(such as neutrino mixing angles) depend in general strongly on the details
of the texture \rfn{texture}, as well as on the form of the right-handed 
neutrino mass matrix. However, the renormalization-induced observables 
are insensitive to these details, being smooth functions of the 
parameters in \rfn{texture} and the heavy neutrino mass eigenvalues only.
This is because the seesaw mechanism 
$m_\nu\sim Y^T_\nu {\cal M}^{-1} Y_\nu$
and the equations \rfn{llrge2}, \rfn{llrge3} depend on the neutrino
parameters in completely different ways.
Therefore our phenomenological analyses of LFV and EDMs is 
only weakly dependent on the exact effective neutrino mixing parameters 
and is quite general for this type of texture models.

We also note that, because the textures are very hierarchical, 
the renormalization effects do not spoil their generic structure.
However, if fine tunings of the parameters in 
\rfn{texture} are required in order to achieve the correct light neutrino
mixing pattern at low energies, these fine tunings 
may not survive the renormalization. Again, our results on the LFV
and EDMs are independent of these fine tunings and are therefore
general.

\begin{figure}[t]
\centerline{\epsfxsize = 0.5\textwidth \epsffile{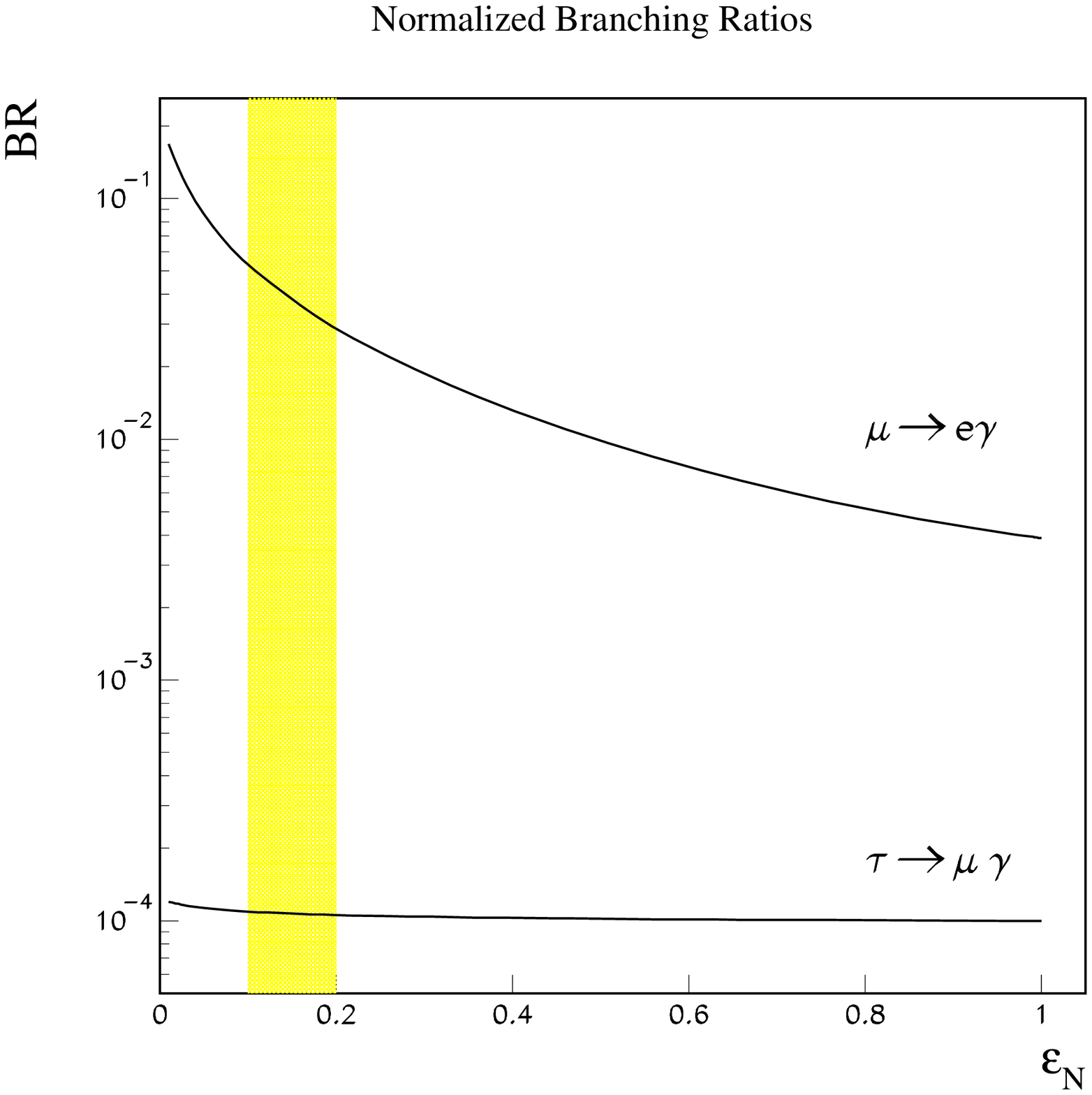} 
\hfill \epsfxsize = 0.5\textwidth \epsffile{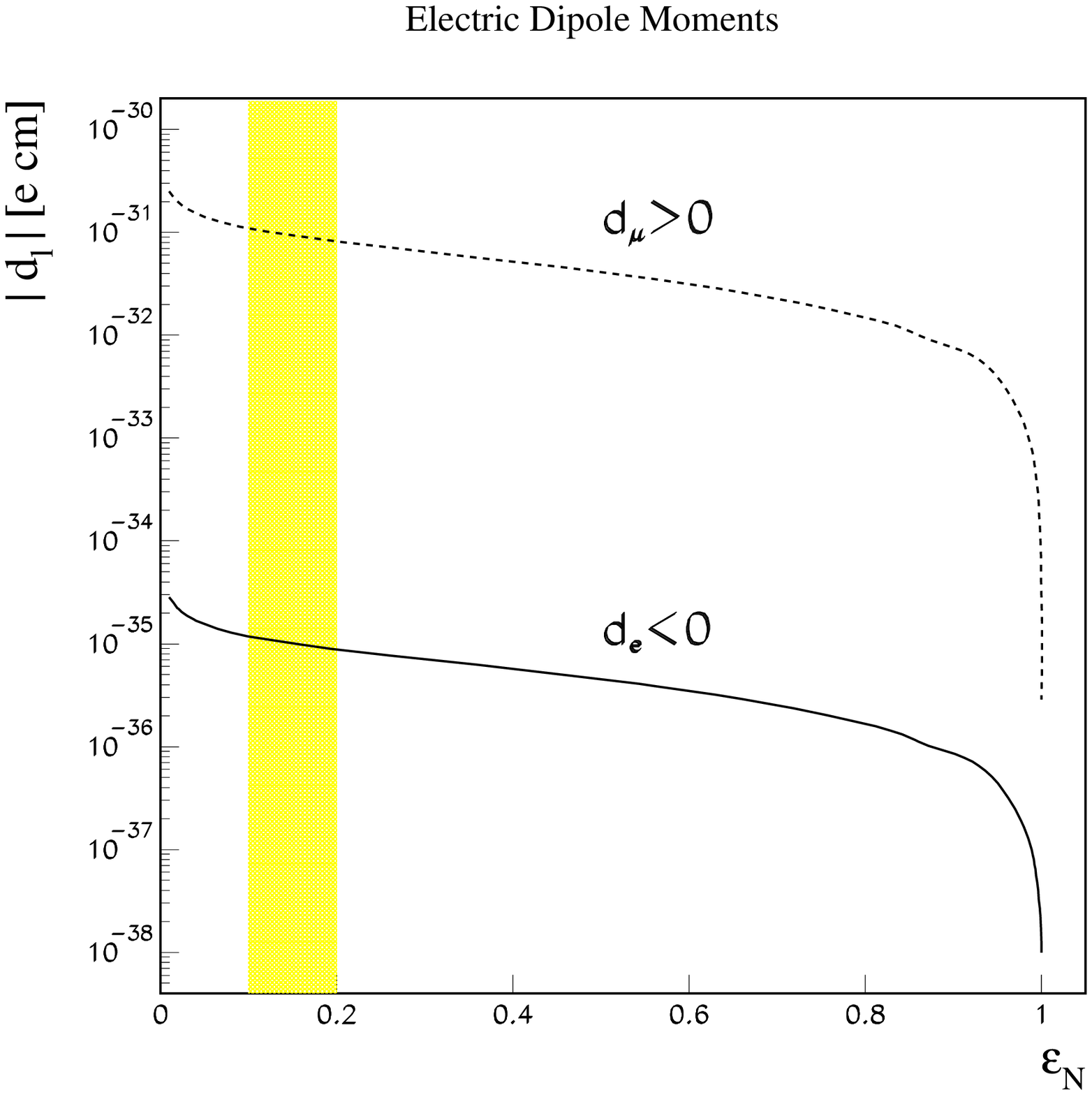} }
\caption{\it Branching ratios for $\mu\to e\gamma$ and $\tau\to\mu\gamma$ 
normalized to their present experimental bounds, and the electron 
and muon electric dipole moments as functions
of the heavy neutrino non-degeneracy parameter $\varepsilon_N.$
Texture parameter values are given in the text, and the favoured range 
of $\varepsilon_N$ is indicated by a vertical shaded band.
\vspace*{0.5cm}}
\label{fig1}
\end{figure}

We first choose 
$Y_0=1,$ $a=1 e^{-i\pi/3},$ $b=1 e^{i 0.3 \pi},$ 
$c=d=1 e^{i\pi/2}$ and $\psi=\pi/2$ in \rfn{texture}.
Since \rfn{texture} predicts hierarchical neutrinos, we must 
have~\footnote{Strictly speaking, this is true if $\varepsilon_\nu$ 
determines the hierarchy in the effective low-energy neutrino masses,
as happens in the models considered.}
$\varepsilon_\nu\sim \sqrt{\Delta m^2_{sol}/\Delta m^2_{atm}}.$
For the currently favoured large-mixing-angle solution to the solar-neutrino
problem, the present oscillation data gives
 at 95\% C.L. the range
$\Delta m^2_{sol}=(2\times 10^{-5}-5\times 10^{-4})$ eV$^2$, 
while for the atmospheric neutrinos one has  
$\Delta m^2_{atm}=(1.4\times 10^{-3}-6\times 10^{-3})$ eV$^2$,
implying phenomenologically $0.6\gsim\varepsilon_\nu\gsim 0.06$.
In our numerical examples we fix $\varepsilon_\nu=0.2$ 
as in \cite{lms}. 
The seesaw fixes heavy neutrino masses in terms of $M_{N_3}=5\times
10^{14}$ GeV.
Taking $\tan\beta=10,$ $m_{1/2}=300$ GeV, $m_{0}=100$ GeV,  $A_{0}=0$
and $sign(\mu)=+1,$ we plot in Fig. \ref{fig1} the branching rations of 
$\mu\to e\gamma$ and $\tau\to\mu\gamma$ normalized to the present experimental
bounds, and the electron and muon
EDMs as functions of the heavy neutrino non-degeneracy parameter
$\varepsilon_N.$ Whilst $Br(\tau\to\mu\gamma)$ depends weakly on 
$\varepsilon_N,$ $Br(\mu\to e\gamma)$ may be increased by more than an
order of magnitude for hierarchical heavy neutrinos.
The dependence of $Br(\mu\to e\gamma)$ on $\varepsilon_N$ is smooth.
We note that for this choice of parameters 
$(g_\mu-2)=4.4\times 10^{-9}$ in a good agreement with the recent
measurement \cite{muedm}.

However, the EDMs show very strong dependences on  $\varepsilon_N.$
For $\varepsilon_N=1$ we always find that $d_\mu/d_e\approx -m_\mu/m_e$
is satisfied with good accuracy. But, already for very small deviations
from unity, the EDMs show very sharp step-function-like increases.
These are typically followed by a steady increase as $\varepsilon_N$
decreases. We note that the signs of $d_e,$ $d_\mu$ and $d_\mu/d_e$ are
not fixed,
and vary depending on the phases in \rfn{texture}. For the given example, 
$d_\mu$ may be enhanced by five and $d_e$ by
four orders of magnitude compared to the degenerate heavy neutrino 
case~\footnote{If one takes $d=0$ in \rfn{texture}, the electron EDM
is reduced by less than an order of magnitude, and the effect on the other 
observables is completely negligible.}.
Moreover, as noted by the shaded vertical bands in the Fig.~\ref{fig1},
the range of $\varepsilon_N$ suggested by neutrino mixing phenomenology
favours relatively large values of the EDMs.

\begin{figure}[t]
\centerline{\epsfxsize = 0.5\textwidth \epsffile{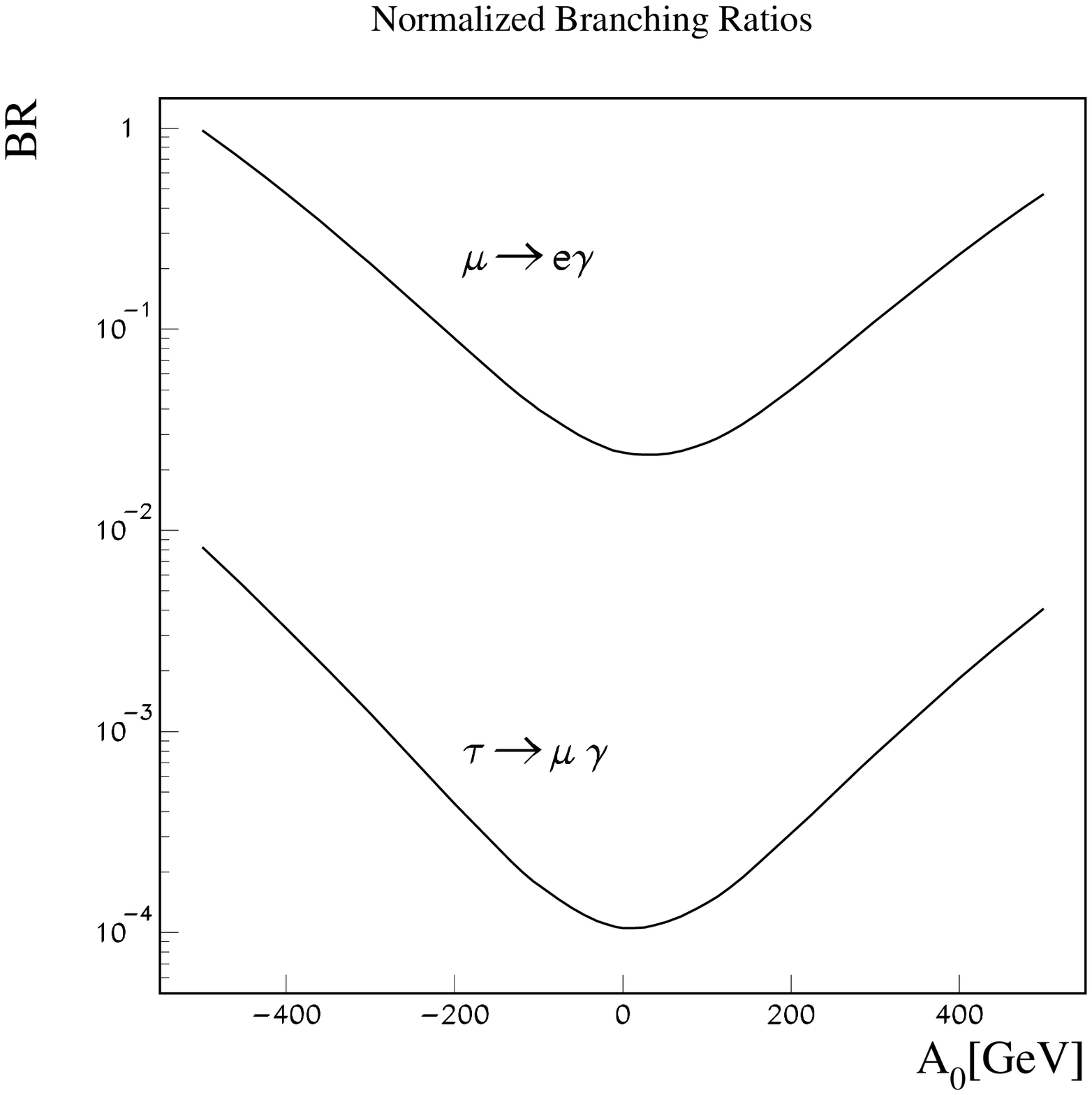} 
\hfill \epsfxsize = 0.5\textwidth \epsffile{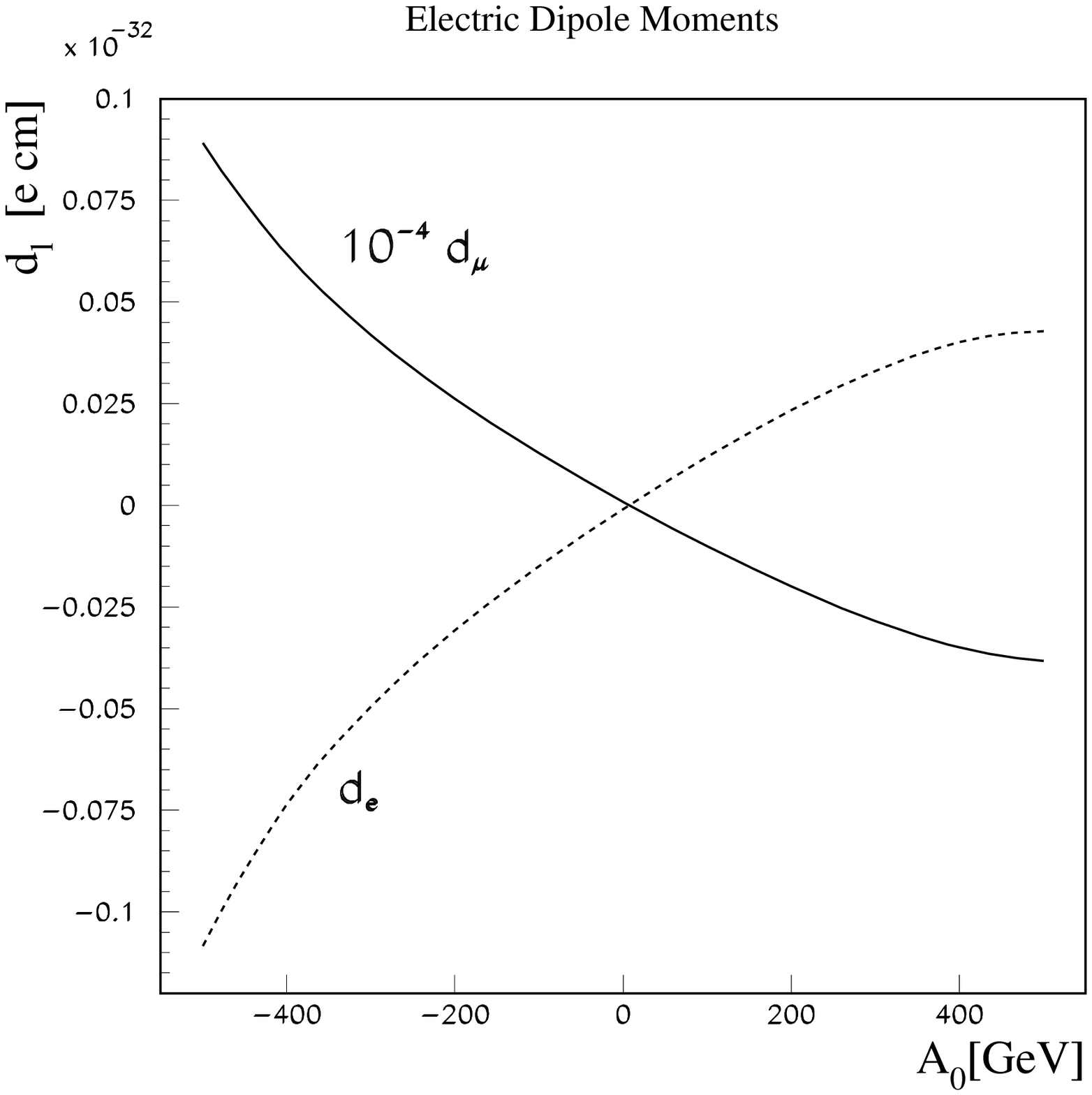} }
\caption{\it Normalized branching ratios and EDMs as functions
of the trilinear soft mass $A_0$ for $\varepsilon_N=0.2.$ 
The rest of the parameters are the same as in
Fig. \ref{fig1}. \vspace*{0.5cm}}
\label{fig2}
\end{figure}

We emphasize that, in the texture models considered, values of
$\varepsilon_N \sim 1$ do not give correct neutrino masses and
mixings. Therefore, our discussion of that region just exemplifies
the importance of the heavy neutrino non-degeneracy effects on EDMs. 
The physically 
meaningful region in Fig. \ref{fig1} is at small values of $\varepsilon_N$
where the EDMs are maximised. In the model \cite{bw} the natural value is
$\varepsilon_\nu\approx\varepsilon_N\sim 0.2.$ 
However, the model \cite{ross2} predicts an even stronger heavy neutrino 
hierarchy: $\varepsilon_N^4:\varepsilon_N^3:1$, with 
$\varepsilon_N\sim\varepsilon_\nu^2$, whilst the texture for the neutrino
Dirac Yukawa matrix is the same. Because the hierarchy between $N_3$
and $N_2$ dominates in the EDMs, the appropriate range is
$\varepsilon_N\sim 0.1$. The vertical band of shading
in Fig. \ref{fig1} indicates the range $0.1 \lsim \varepsilon_N \lsim 0.2$ 
favoured in our sampling of models.

The dependence of the LFV processes and EDMs on the SUSY soft masses
$m_{1/2}$ and $m_{0}$ enters mainly through sparticle propagators and
is not dramatic. However, as we have shown before, the dependence
of EDMs on $A_0$ is non-trivial. This is shown in Fig. \ref{fig2}
where we plot the normalized branching ratios and EDMs
as functions of $A_0.$ Notice that $d_\mu$ is scaled by $10^{-4}$
to fit the figure. We have fixed $\varepsilon_N=0.2$ and chosen
all the other parameters as in Fig. \ref{fig1}. Comparison with  Fig.
\ref{fig1}
shows that, for $A_0\neq 0$, an additional two orders of magnitude 
enhancement of the EDMs is possible compared to the $A_0= 0$ case.
Also the signs of the EDMs depend on the sign of $A_0.$

\begin{figure}[t]
\centerline{\epsfxsize = 0.5\textwidth \epsffile{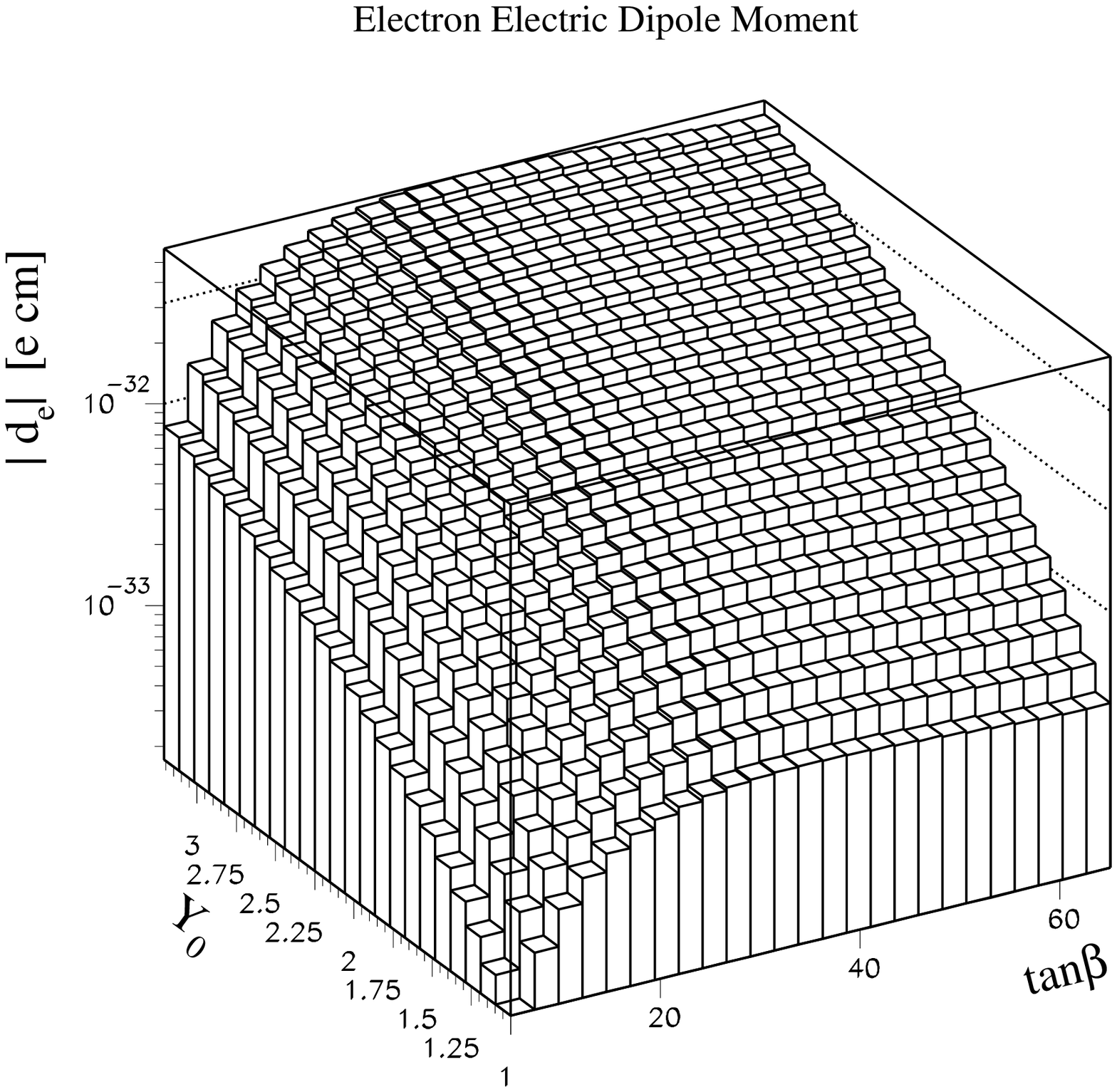} 
\hfill \epsfxsize = 0.5\textwidth \epsffile{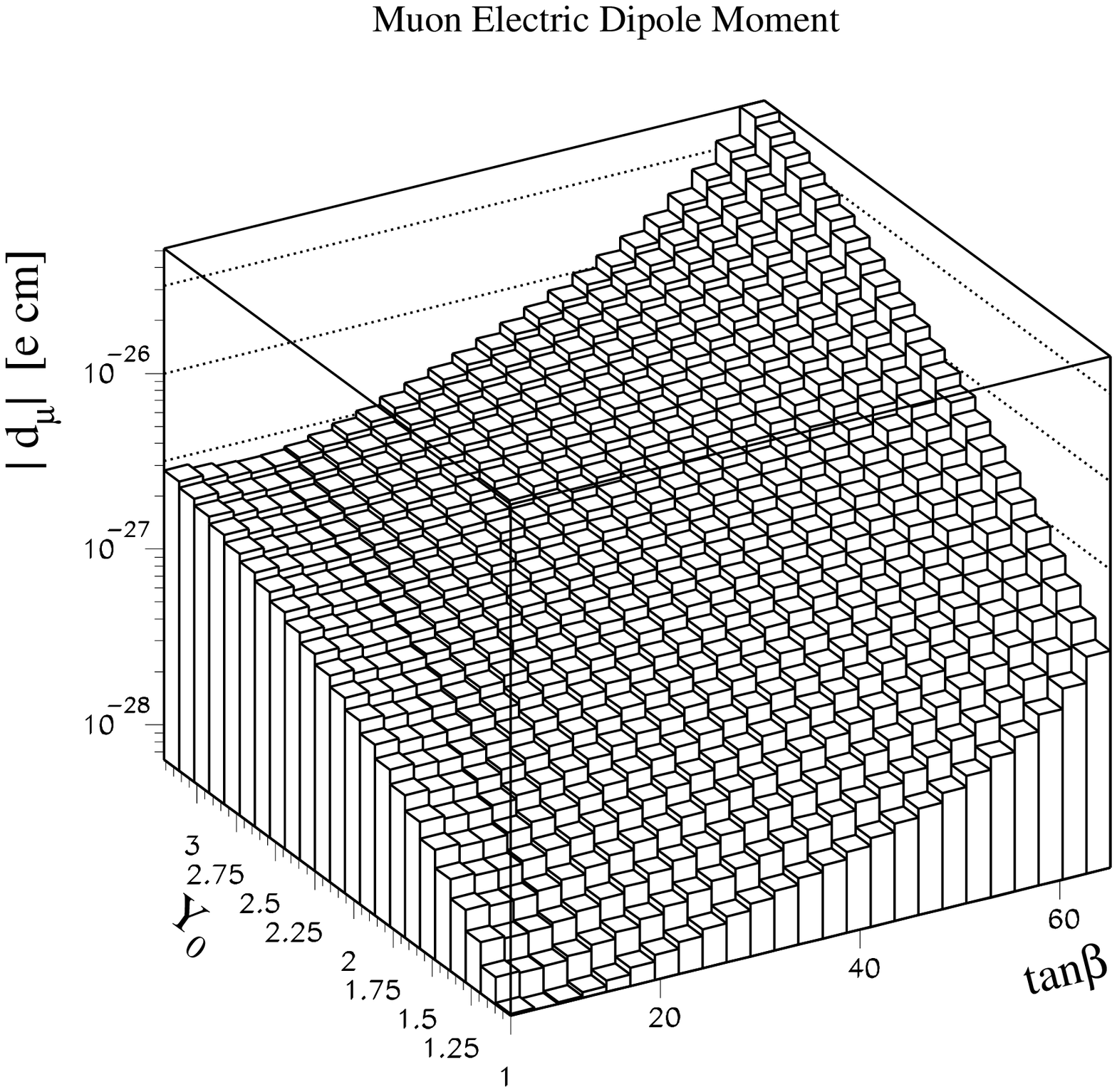} }
\caption{\it   Absolute values of EDMs as functions of 
$\tan\beta$ and the overall Yukawa scale factor $Y_0.$ 
Other parameter values are given in the text. 
\vspace*{0.5cm}}
\label{fig3}
\end{figure}

How large EDMs can one obtain in those texture models? We take 
$a=3 e^{i 0.73\pi},$ $b=3 e^{i\pi/5},$ 
$c=d=0.05 e^{i\pi/2},$ $\psi=\pi/2$ and plot in  Fig. \ref{fig3}
the absolute values of the electron and muon EDMs as functions of 
$\tan\beta$ and the overall Yukawa scale factor $Y_0.$ 
For the given choice of phases both $d_e$ and $d_\mu$ are negative.
The branching ratios of $\mu\to e\gamma$ and $\tau\to\mu\gamma$ are
below the present experimental bounds over the whole considered
parameter space. The heavy neutrino mass $M_{N_3}$ is calculated
from the seesaw mechanism taking the heaviest light neutrino mass to be
$m_{\nu_3}=0.06$ eV, and the neutrino mass hierarchy is fixed by
$\varepsilon_N=0.1.$ While increasing $\tan\beta$ from $5$ to $65,$
we also increase the soft mass term  $m_{0}$ linearly from 100 GeV to 
700 GeV, and decrease $A_{0}$ from -500 GeV to -700 GeV.
The gaugino mass is fixed to be $m_{1/2}=400$ GeV.
Note that, for $sign(\mu)=+1$,
the supersymmetric radiative corrections reduce the $b$-quark Yukawa
coupling by $\sim 50\%$ and its running is under control even for such 
a large $\tan\beta$ as 65.

We see in  Fig. \ref{fig3} that $|d_e|$  and $|d_\mu|$ depend quite 
strongly on the size of the Yukawa couplings as well as on $\tan\beta.$
While the former dependence is trivially expected, the latter is 
particular to the considered texture \rfn{texture}. Because this
texture is hierarchical with small off-diagonal elements, the 
renormalization-induced EDMs depend on the size of the charged-lepton 
Yukawa matrix $Y_e$ 
which is maximized for large $\tan\beta.$ For different textures the
$\tan\beta$ behaviour might be different.

We point out that for $Y_0\sim 3$ the largest Yukawa coupling 
$(Y_\nu)_{33}$ is almost at the fixed point. It is interesting that the
Super-Kamiokande measurement of light neutrino masses, the seesaw mechanism
and the fixed-point idea for up-type Yukawas are perfectly consistent 
with each other. Starting with some large value for the Yukawa coupling at
$M_{GUT},$ at the scale $M_{N_3}\sim 5\times 10^{15}$ GeV where the 
heaviest singlet neutrino decouples, one always gets 
$(Y_\nu)_{33}\sim 3$, which implies just the correct effective neutrino 
mass for Super-Kamiokande.  Thus the fixed point idea would 
suggest large values for EDMs.
Therefore, $|d_\mu|$ may exceed the $10^{-26}$ e cm level,
and  $|d_e|$ exceeds $10^{-33}$ e cm for almost the entire parameter
space considered. These values for $|d_\mu|$ and  $|d_e|$ are in the range of 
interest for future experiments \cite{PRISM,nf,edmnew}.

We stress that our analyses of EDMs using textures is
not phenomenologically general nor exhaustive. It just illustrates
what can happen in realistic models. In more general cases
(choosing purely phenomenological $Y_\nu$, relaxing the exact
universality conditions on the soft supersymmetry-breaking terms, etc.)
somewhat different results can be expected. In particular,
large EDMs may appear even when $\tan\beta$ is not large.
For example, working with the Yukawa texture predicted by the 
$U(2)$ flavour model~\cite{bcr}, large EDMs can be achieved for small
values of $\tan\beta$ if some hierarchy between $M_{N_3}$ and $M_{N_2}$ 
is allowed. This is because this texture is
asymmetric with large $(33)$ and $(23)$ elements.

In conclusion: we have shown that, in the minimal supersymmetric seesaw 
model with non-degenerate heavy neutrino masses, the charged-lepton EDMs 
may be enhanced by several orders of magnitude compared to the
degenerate heavy-neutrino case. This is because, in the leading 
logarithmic approximation,
three additional  physical phases in $Y_\nu$ renormalize the soft 
supersymmetry-breaking mass terms. One of these phases contributes also to 
leptogenesis. We show that, within a plausible class of Yukawa texture
models, the charged-lepton EDMs may reach values testable in proposed 
experiments, in agreement with general analyses \cite{konstantin}.
In more general cases, we expect the allowed values of EDMs to increase 
further. Combining the measurements of EDMs with all possible neutrino and
charged-lepton-flavour- and CP-violating measurements would thus allow
us to obtain information on the leptogenesis mechanism from 
low-energy experiments.

\vskip 0.5in
\vbox{
\noindent{ {\bf Acknowledgements} } \\
\noindent  
This work is partially supported by EU TMR
contract No.  HPMF-CT-2000-00460, ESF grant No. 3832, and the
Grant-in-Aid for Scientific Research from the Ministry of Education,
Science, Sports and Culture of Japan (Priority Area 707
`Supersymmetry and Unified Theory of Elementary Particles',
No. 13001292).

}


\begin{thebibliography}{99}


\bibitem{seesaw} M. Gell-Mann, P. Ramond and R. Slansky, Proceedings of   
the Supergravity Stony Brook Workshop, New York, 1979, eds. P. Van   
Nieuwenhuizen and D. Freedman (North-Holland, Amsterdam);
T. Yanagida, Proceedings of
the  Workshop  on Unified  Theories  and  Baryon  Number in the  
Universe,  Tsukuba,  Japan 1979 (edited by A.  Sawada and A.
Sugamoto, KEK Report No.  79-18, Tsukuba); 
R.~Mohapatra and G.~Senjanovic, 
Phys.\ Rev.\ Lett.\ {\bf 44} (1980) 912.

\bibitem{masiero}
For a general analyses, see, \eg,
F.~Gabbiani, E.~Gabrielli, A.~Masiero and L.~Silvestrini,
Nucl.\ Phys.\ B {\bf 477} (1996) 321.

\bibitem{bm}
F.~Borzumati and A.~Masiero,
Phys.\ Rev.\ Lett.\ {\bf 57} (1986) 961.

\bibitem{Blazek:2001zm}
T.~Blazek and S.~F.~King,
Phys.\ Lett.\ B {\bf 518} (2001) 109.

\bibitem{review}
For a review and references see, \eg,
Y. Kuno and  Y. Okada, 
Rev.\ Mod.\ Phys.\ {\bf 73} (2001) 151.

\bibitem{h1}
J.~Hisano, T.~Moroi, K.~Tobe and M.~Yamaguchi,
Phys.\ Rev.\ D {\bf 53} (1996) 2442;
J.~Hisano and D.~Nomura,
Phys.\ Rev.\ D {\bf 59} (1999) 116005.

\bibitem{Okada} Y.~Okada, K.~Okumura and Y.~Shimizu,
Phys.\ Rev.\ D {\bf 58} (1998) 051901;
Phys.\ Rev.\ D {\bf 61} (2000) 094001.


\bibitem{strumia}
A.~Romanino and A.~Strumia,
hep-ph/0108275; and references therein.

\bibitem{ross}
R. G. Roberts, A. Romanino, G. G. Ross and L. Velasco-Sevilla,
hep-ph/0104088; and references therein.

\bibitem{lms}
S. Lavignac, I. Masina and C. A. Savoy, hep-ph/0106245.

\bibitem{ross2}
S. F. King and G. G. Ross, hep-ph/0108112. 

\bibitem{bw}
W. Buchm\"uller and D. Wyler,
hep-ph/0108216.

\bibitem{lepto}
M.~Fukugita and T.~Yanagida,
Phys.\ Lett.\ B {\bf 174} (1986) 45.

\bibitem{so10lepto}
W.~Buchm\"uller and M.~Pl\"umacher,
Phys.\ Lett.\ B {\bf 389} (1996) 73;
M.~Pl\"umacher, Phys.\ Lett.\ {\bf 530} (1998) 207;
W.~Buchm\"uller and T.~Yanagida,
Phys.\ Lett.\ B {\bf 445} (1999) 399.


\bibitem{mnbuedm}
Y.~K.~Semertzidis {\it et al.},
hep-ph/0012087.

\bibitem{PRISM}
M. Furusaka {\it et al.}, JAERI/KEK Joint Project Proposal 
{\it The Joint Project for High-Intensity Proton Accelerators},
KEK-REPORT-99-4, JAERI-TECH-99-056.


\bibitem{nf}
J. \"Ayst\"o {\it et al.}, {\it Physics with Low-Energy Muons at a 
Neutrino Factory Complex}, CERN-TH/2001-231, hep-ph/0109217.

\bibitem{edmnew}
S. K. Lamoreaux, nucl-ex/0109014.

\bibitem{Barbieri:1995tw}
R.~Barbieri, L.~J.~Hall and A.~Strumia,
Nucl.\ Phys.\ B {\bf 445} (1995) 219.

\bibitem{Arkani-Hamed:1995fs}
N.~Arkani-Hamed, H.~C.~Cheng and L.~J.~Hall,
Phys.\ Rev.\ D {\bf 53} (1996) 413;
J.~Hisano, D.~Nomura, Y.~Okada, Y.~Shimizu and M.~Tanaka,
Phys.\ Rev.\ D {\bf 58} (1998) 116010.

\bibitem{Babu:2000dq}
K.~S.~Babu, B.~Dutta and R.~N.~Mohapatra,
Phys.\ Rev.\ Lett.\  {\bf 85} (2000) 5064.

\bibitem{cpt}
For a  reference list see, \eg, 
T.~Ota, J.~Sato and Y.~Kuno, 
hep-ph/0107007.

\bibitem{ehlr}
J. Ellis, J. Hisano, S. Lola and M. Raidal, hep-ph/0109125.

\bibitem{acfh}
N.~Arkani-Hamed, J.~L.~Feng, L.~J.~Hall and H.~Cheng,
Nucl.\ Phys.\ B {\bf 505} (1997) 3.

\bibitem{valle}
For an early analysis of phase counting, see
J.~Schechter and J.~W.~Valle,
Phys.\ Rev.\ D {\bf 22} (1980) 2227;
Phys.\ Rev.\ D {\bf 23} (1981) 1666.


\bibitem{khalil}
S.~Abel, S.~Khalil and O.~Lebedev,
Nucl.\ Phys.\ B {\bf 606} (2001) 151.

\bibitem{nedm}
S.~Bertolini and F.~Vissani,
Phys.\ Lett.\ B {\bf 324} (1994) 164;
T.~Inui, Y.~Mimura, N.~Sakai and T.~Sasaki,
Nucl.\ Phys.\ B {\bf 449} (1995) 49;
S.~A.~Abel, W.~N.~Cottingham and I.~B.~Whittingham,
Phys.\ Lett.\ B {\bf 370} (1996) 106;
A.~Romanino and A.~Strumia,
Nucl.\ Phys.\ B {\bf 490} (1997) 3;
C.~Hamzaoui, M.~Pospelov and R.~Roiban,
Phys.\ Rev.\ D {\bf 56} (1997) 4295.

\bibitem{vissani}
L.~Covi, E.~Roulet and F.~Vissani,
Phys.\ Lett.\ B {\bf 384} (1996) 169;
W.~Buchm\"uller and M.~Pl\"umacher,
Phys.\ Rept.\  {\bf 320} (1999) 329;
M.~Flanz, E.~A.~Paschos, U.~Sarkar and J.~Weiss,
Phys.\ Lett.\ B {\bf 389} (1996) 693;
A.~Pilaftsis,
Phys.\ Rev.\ D {\bf 56} (1997) 5431,
Int.\ J.\ Mod.\ Phys.\ A {\bf 14} (1999) 1811.


\bibitem{nath}
T.~Ibrahim and P.~Nath,
Phys.\ Rev.\ D {\bf 57} (1998) 478
[Erratum - {\it ibid.} {\bf 58} (1998) 019901].

\bibitem{mega}
M. L. Brooks \ea, [MEGA Collaboration],
\prl{83} (1999) 1521.

\bibitem{cleo}
S. Ahmed \ea, [CLEO Collaboration],
Phys.\ Rev.\ D {\bf 61} (2000) 071101.

\bibitem{PSI}
L.M.~Barkov {\it et al.}, Research Proposal for experiment at PSI (1999).

\bibitem{eedm}
E.~D.~Commins, S.~B.~Ross, D.~DeMille and B.~C.~Regan,
Phys.\ Rev.\ A {\bf 50} (1994) 2960.

\bibitem{muedm}
H.~N.~Brown {\it et al.}  [Muon g-2 Collaboration],
Phys.\ Rev.\ Lett.\  {\bf 86} (2001) 2227.

\bibitem{tauedm}
M.~Acciarri {\it et al.}  [L3 Collaboration],
Phys.\ Lett.\ B {\bf 434} (1998) 169.

\bibitem{mv}
S. P. Martin and M. T. Vaughn, Phys.\ Rev.\ D {\bf 50} (1994) 2282.

\bibitem{bcr}
R. Barbieri, P. Creminelli and A. Romanino, 
Nucl.\ Phys.\ B {\bf 559} (1999) 17.

\bibitem{konstantin}
T.~Ibrahim and P.~Nath, 
hep-ph/0105025;
J.~L.~Feng, K.~T.~Matchev and Y.~Shadmi,
hep-ph/0107182 and hep-ph/0110157.


\end{thebibliography}
\end{document}